\documentclass[3p,times]{elsarticle}  
\usepackage{graphicx} 
\usepackage{amssymb} 
\usepackage{amsmath} 
\usepackage{amsfonts} 
\usepackage{epsfig} 
\journal{Physics Letters B} 

\begin{document} 

\begin{frontmatter} 

\title{The $d^{\ast}$(2380) dibaryon resonance width and 
decay branching ratios} 
\author{A.~Gal\corref{cor1}} 
\cortext[cor1]{corresponding author: Avraham Gal, avragal@savion.huji.ac.il}  
\address{Racah Institute of Physics, The Hebrew University, Jerusalem 91904, 
Israel} 

\begin{abstract} 
Attempts to reproduce theoretically the width $\Gamma_{d^{\ast}}=80\pm 10
$~MeV of the $I(J^P)=0(3^+)$ $d^{\ast}$(2380) dibaryon resonance established 
by the WASA-at-COSY Collaboration are discussed. The validity of associating 
the $d^{\ast}$(2380) in quark-based models exclusively with a tightly bound 
$\Delta\Delta$ configuration is questioned. The $d^{\ast}$(2380) width and 
decay branching ratios into $NN\pi\pi$, $NN\pi$ and $NN$ final states are 
studied within the Gal-Garcilazo hadronic model in which the $d^{\ast}$(2380) 
is a $\pi N\Delta$ resonance embedded in the $NN\pi\pi$ continuum some 80~MeV 
below the $\Delta\Delta$ threshold. In particular, predictions are made for 
the branching ratios of the unobserved yet $d^{\ast}(2380)\to NN\pi$ decays 
which are suppressed in a purely-$\Delta\Delta$ dibaryon model. Comments are 
also made on a possible connection of the ABC effect observed in the $pn\to 
d^{\ast}\to d\pi^0\pi^0$ resonance reaction to the $d^{\ast}$(2380) dibaryon. 
\end{abstract} 


\end{frontmatter}

\section{Introduction} 
\label{sec:intro} 

\begin{figure}[!ht] 
\begin{center} 
\includegraphics[width=0.48\textwidth,height=5cm]{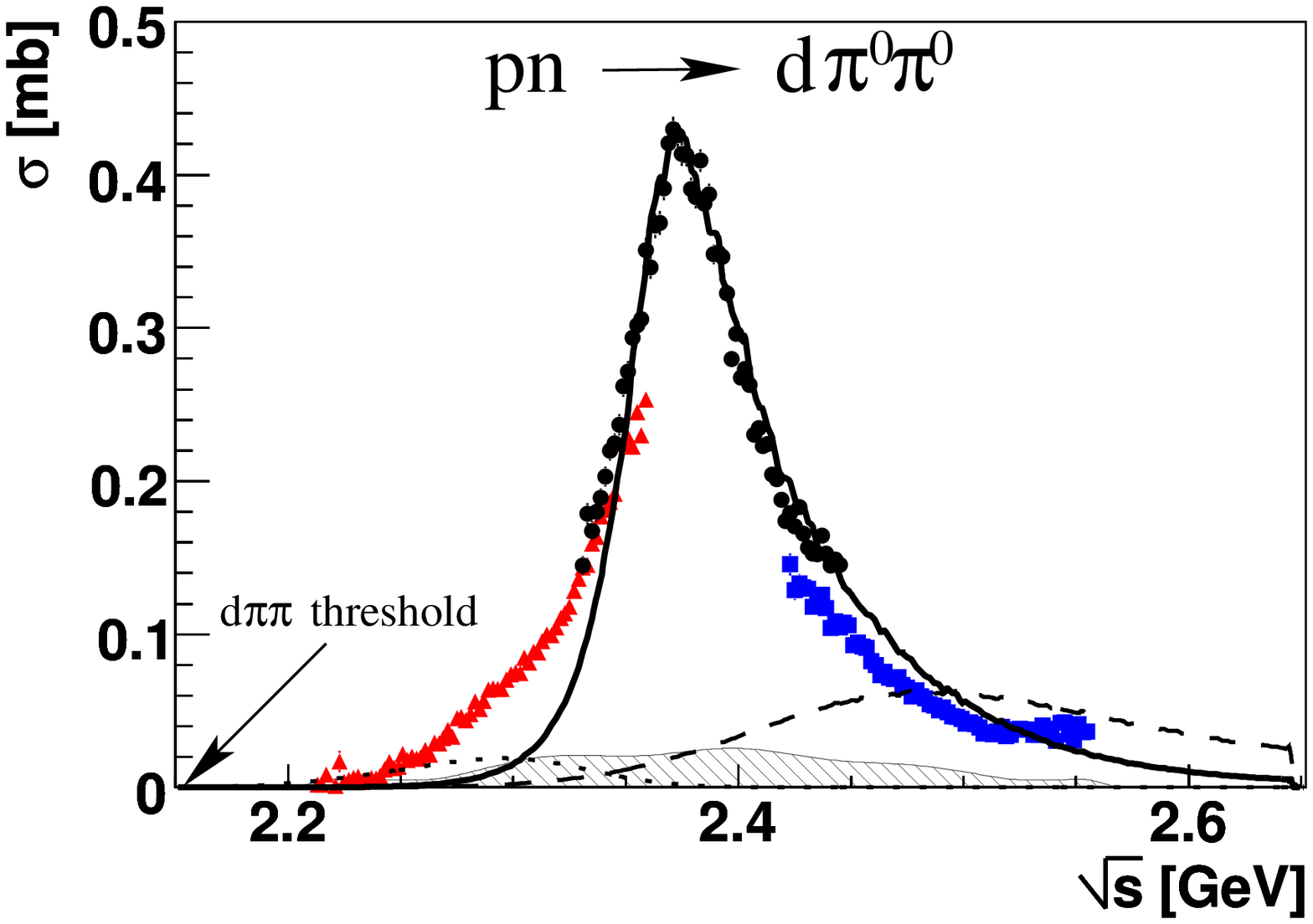} 
\includegraphics[width=0.48\textwidth,height=5cm]{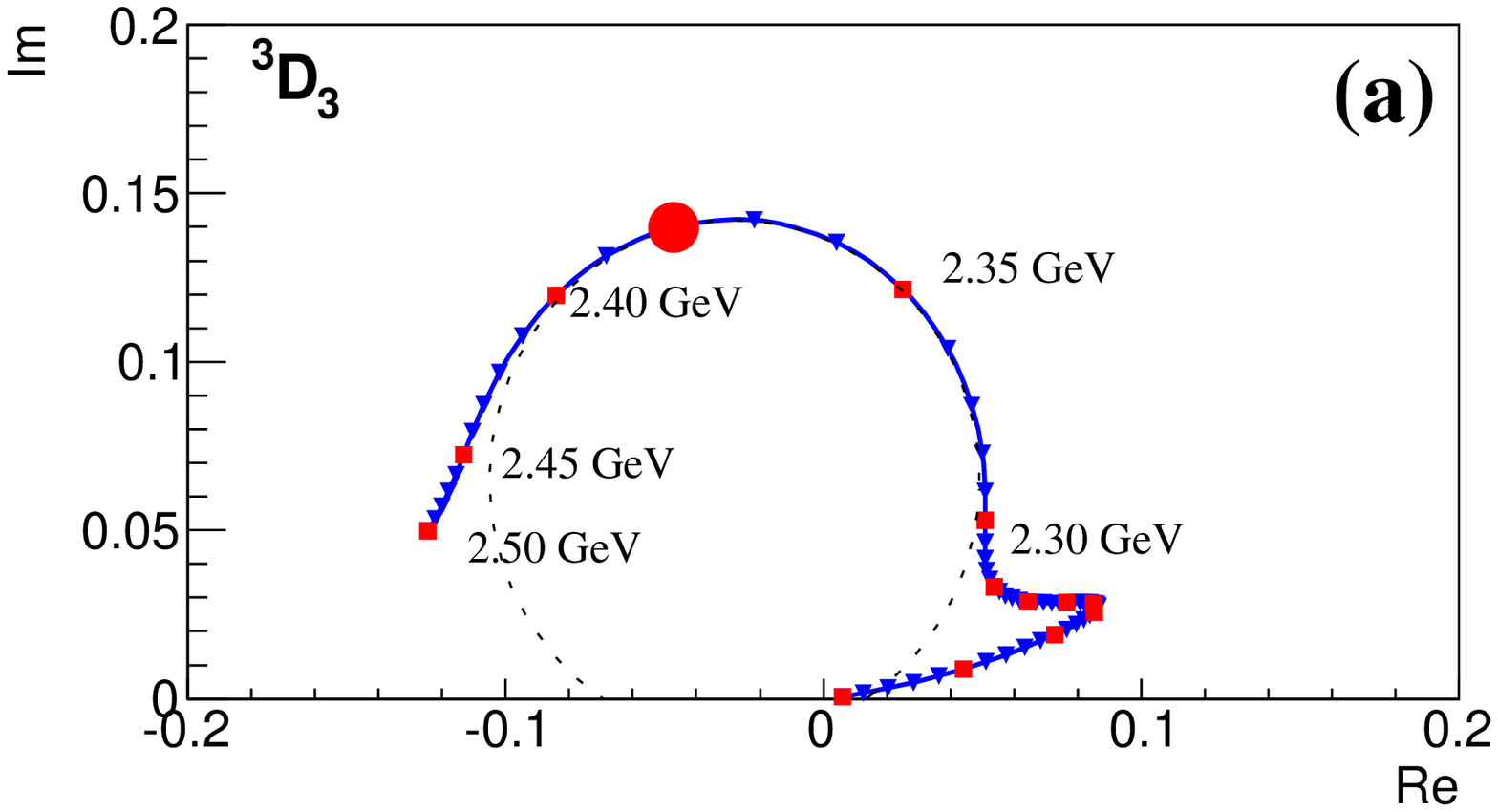} 
\caption{$d^{\ast}$(2380) dibaryon resonance signatures in recent 
WASA-at-COSY Collaboration experiments. Left: from the peak observed in 
the $pn\to d\pi^0\pi^0$ reaction \cite{wasa11}. Right: from the Argand 
diagram of the $^3D_3$ partial wave in $pn$ scattering \cite{wasa14}.} 
\label{fig:WASA} 
\end{center} 
\end{figure} 

The WASA-at-COSY Collaboration observed a relatively narrow peak, 
$\Gamma_{d^{\ast}}\approx 70$~MeV, about 80~MeV below the $\Delta\Delta$ 
threshold in the $pn\to d\pi^0\pi^0$ reaction~\cite{wasa11}. This peak, 
shown on the left panel of Fig.~\ref{fig:WASA}, was identified with the 
$I(J^P)$=$0(3^+)$ ${\cal D}_{03}(2350)$ $\Delta\Delta$ dibaryon predicted 
in 1964 by Dyson and Xuong~\cite{dyson64}. The $I=0$ isospin assignment 
follows from the isospin balance in $pn \to d\pi^0\pi^0$, and the $J^P=3^+$ 
spin-parity assignment follows from the measured deuteron angular 
distribution. The $d^{\ast}$(2380) was also observed in 
$pn\to d\pi^+\pi^-$, with cross section consistent with that 
measured in $pn\to d\pi^0\pi^0$~\cite{wasa13a}, and studied in several 
other related $pn\to NN\pi\pi$ reactions~\cite{wasa13b,wasa13c,hades15}. 
Recent measurements of $pn$ scattering and analyzing power~\cite{wasa14} 
have led to the $pn$ $^3D_3$ partial-wave Argand diagram shown on the right 
panel of Fig.~\ref{fig:WASA}, supporting the $d^{\ast}$(2380) dibaryon 
resonance interpretation. 

The mass of a possible $I(J^P)$=$0(3^+)$ $\Delta\Delta$ dibaryon has been 
the subject of many quark-based calculations~\cite{clement17} but its width 
received little attention, and that~\cite{huang14,dong16} only since the 
discovery of the $d^{\ast}$(2380). The term `quark-based' does not necessarily 
mean that the resulting $d^{\ast}$(2380) is of a purely {\it hexaquark} 
structure. In fact, a recent quark-model study of spatially symmetric $L=0$ 
$6q$ states finds the $I(J^P)$=$0(3^+)$ hexaquark several hundreds of MeV 
above the $\Delta\Delta$ threshold~\cite{PPL15}. It is by adding potentially 
double-counting meson exchanges, e.g. a scalar-isoscalar $\sigma$ meson, 
and applying resonating group methods (RGM), that quark-based calculations 
generate a tightly bound and compact $\Delta\Delta$ dibaryon. 

The $d^{\ast}$(2380) was also studied recently~\cite{galgar13,galgar14} within 
a $\pi{\cal D}_{12}$--$\Delta\Delta$ coupled-channels $\pi N\Delta$ hadronic 
model, using $\pi N$ and $N\Delta$ pairwise interactions each of which 
produces its own resonance: the $I(J^P)$=$\frac{3}{2}({\frac{3}{2}}^+)$ 
$\Delta$(1232) baryon, and the $I(J^P)$=1(2$^+$) ${\cal D}_{12}$(2150) 
dibaryon resonance generated by solving $NN\pi$ three-body Faddeev equations. 
The $d^{\ast}$(2380) $S$-matrix pole in this model is embedded in the 
$NN\pi\pi$ continuum, about midway between the corresponding two-body 
thresholds, giving rise to a two-component structure: a resonance with respect 
to the lower $\pi{\cal D}_{12}$ threshold and a tightly bound state with 
respect to the upper $\Delta\Delta$ threshold. This coupled-channels structure 
of the $d^{\ast}$(2380) dibaryon is absent in quark-based $\Delta\Delta$ 
dibaryon models. 

In this note, we discuss the role of the lower channel $\pi{\cal D}_{12}$ 
in explaining the $d^{\ast}$(2380) width $\Gamma_{d^{\ast}}\approx 70$~MeV 
(left panel of Fig.~\ref{fig:WASA}) which is considerably smaller than twice 
the width of a single $\Delta$ baryon, $\Gamma_{\Delta}\approx 115$~MeV. 
It is shown in the next section that the $d^{\ast}$(2380) width would have 
been even smaller than its observed value, were it not restrained by the 
effect of the $\pi{\cal D}_{12}$ channel. In a subsequent section we discuss 
in some detail the $d^{\ast}$(2380) partial decay widths and decay branching 
ratios in comparison to those deduced from experiment~\cite{BCS15}. 
Predictions are made in particular for the $d^{\ast}(2380)\to NN\pi$ 
partial decay widths which are suppressed to leading order within 
a $\Delta\Delta$ single-channel description of the $d^{\ast}$(2380). 
We also comment on a possible connection of the ABC effect observed in 
$pn\to d\pi^0\pi^0$~\cite{BCS17} to the $d^{\ast}$(2380) dibaryon.

\section{Is the $d^{\ast}$(2380) $\Delta\Delta$ dibaryon a compact 
or extended object?} 
\label{sec:compact} 

Assuming a quasibound $\Delta\Delta$ configuration for the $d^{\ast}$(2380) 
dibaryon, the phase space for a given $\Delta_j\to N\pi$ decay ($j=1,2$) 
to occur independently of the other decay is reduced by binding: 
$M_{\Delta}=1232 \Rightarrow 1232-B_{\Delta\Delta}/2$ MeV, where $B_{\Delta
\Delta}=2\times 1232-2380=84$~MeV is the binding energy of the two $\Delta$s. 
This reduces the $\Delta$ free-space width, $\Gamma_{\Delta}\approx 115$~MeV 
\cite{SP07,anisovich12}, to 81~MeV using Eq.~(\ref{eq:gamma}) below. 
However, this simple estimate is incomplete, as realized recently also by 
Niskanen~\cite{niskanen16}, since neither of the two $\Delta$s is at rest 
within such a deeply bound $\Delta\Delta$ state. To take account of the 
$\Delta\Delta$ momentum distribution, we evaluate the bound-$\Delta$ decay 
width ${\overline{\Gamma}}_{\Delta\to N\pi}$ by averaging $\Gamma_{\Delta\to 
N\pi}(\sqrt{s_{\Delta}})$ over the $\Delta\Delta$ bound-state momentum-space 
wavefunction squared, 
\begin{equation} 
{\overline{\Gamma}}_{\Delta\to N\pi}\equiv\langle \Psi^{\ast}(p_{\Delta\Delta})
|\Gamma_{\Delta\to N\pi}(\sqrt{s_{\Delta}})|\Psi(p_{\Delta\Delta})\rangle 
\approx \Gamma_{\Delta\to N\pi}(\sqrt{{\overline{s}}_{\Delta}}), 
\label{eq:av} 
\end{equation} 
with $s_{\Delta}$ the invariant energy squared and its average bound-state 
value ${\overline{s}}_{\Delta}$ defined by 
\begin{equation} 
s_{\Delta}=(1232-B_{\Delta\Delta}/2)^2-p_{\Delta\Delta}^2, \,\,\,\,\,\, 
{\overline{s}}_{\Delta}=(1232-B_{\Delta\Delta}/2)^2-P_{\Delta\Delta}^2 , 
\label{eq:s} 
\end{equation} 
in terms of a $\Delta\Delta$ bound-state variable momentum $p_{\Delta\Delta}$ 
and its r.m.s. value $P_{\Delta\Delta}\equiv{\langle p_{\Delta\Delta}^2
\rangle}^{1/2}$. 

\begin{table}[hbt] 
\begin{center} 
\caption{Values of $\sqrt{{\overline{s}}_{\Delta}}$ as a function of 
$R_{\Delta\Delta}$, using $P_{\Delta\Delta}R_{\Delta\Delta}=\frac{3}{2}$ in 
Eq.~(\ref{eq:s}), values of the corresponding decay-pion momentum 
${\overline{q}}_{\Delta\to N\pi}$, values of ${\overline{\Gamma}}_{\Delta\to 
N\pi}$ from Eq.~(\ref{eq:gamma}) and of ${\overline{\Gamma}}_{\Delta\Delta\to 
NN\pi\pi}\simeq\frac{5}{3}{\overline{\Gamma}}_{\Delta\to N\pi}$.} 
\begin{tabular}{ccccc} 
\hline 
$R_{\Delta\Delta}$ (fm) & $\sqrt{{\overline{s}}_{\Delta}}$ (MeV) & 
${\overline{q}}_{\Delta\to N\pi}$ (MeV) & ${\overline{\Gamma}}_{\Delta\to 
N\pi}$ (MeV) & ${\overline{\Gamma}}_{\Delta\Delta\to NN\pi\pi}$ (MeV) \\ 
\hline 
0.6 & 1083 & 38.3  & 1.6  & 2.6  \\ 
0.7 & 1112 & 96.6  & 19.3 & 32.1 \\ 
0.8 & 1131 & 122.0 & 33.5 & 55.8 \\ 
1.0 & 1153 & 147.7 & 50.6 & 84.4  \\
1.5 & 1174 & 170.4 & 67.4 & 112.3  \\
2.0 & 1181 & 177.9 & 73.2 & 122.0  \\  
\hline 
\end{tabular}  
\label{tab:width} 
\end{center} 
\end{table} 

\begin{figure}[htb] 
\begin{center} 
\includegraphics[width=0.4\textwidth]{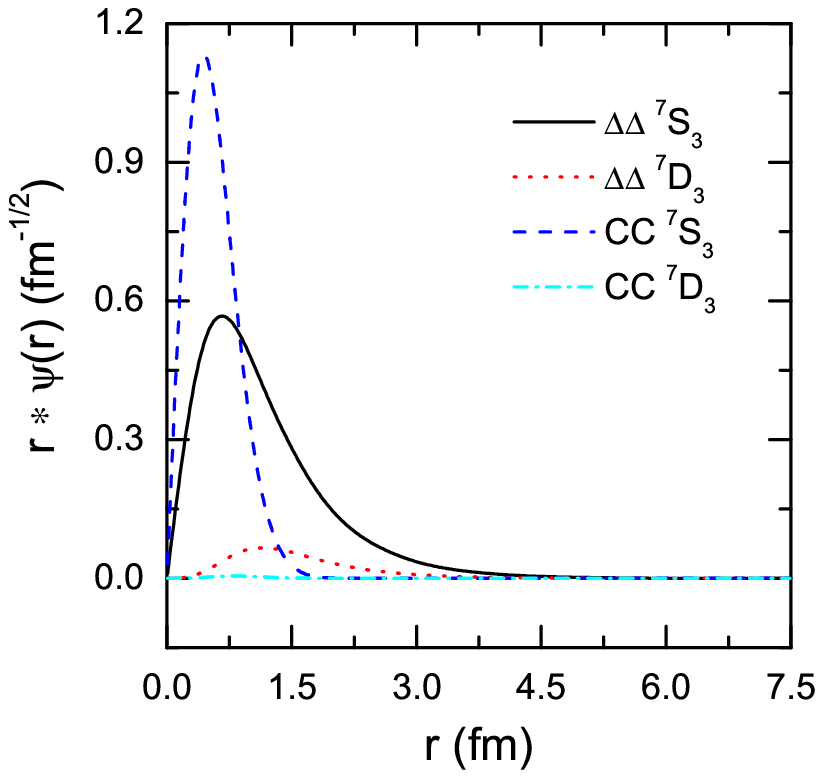} 
\includegraphics[width=0.4\textwidth]{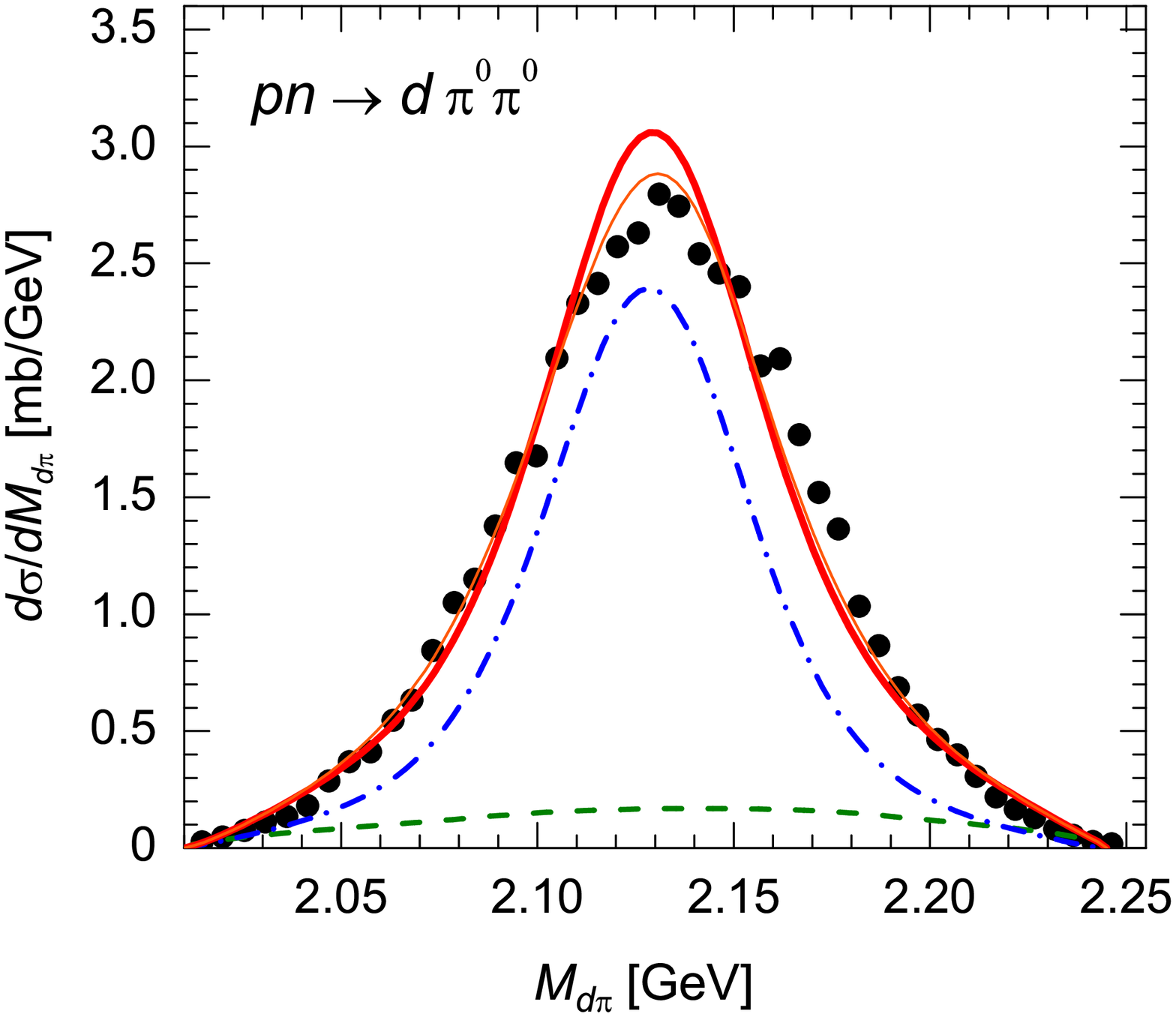} 
\caption{Left: a $d^{\ast}$(2380) $\Delta\Delta$ wavefunction 
with r.m.s. radius $R_{\Delta\Delta}=0.76$~fm from quark-based RGM 
calculations~\cite{huang15}. Right: The $pn\to d\pi^0\pi^0$ WASA-at-COSY 
$M_{d\pi}$ invariant-mass distribution \cite{wasa11} and, in solid lines, 
as calculated \cite{PK16} for two input parametrizations of ${\cal D}_{12}
(2150)$. The dot-dashed line gives the $\pi{\cal D}_{12}(2150)$ contribution 
to the two-body decay of the $d^{\ast}$(2380) dibaryon, and the dashed line 
gives a $\sigma$-meson emission contribution.} 
\label{fig:wf} 
\end{center} 
\end{figure}

In Table~\ref{tab:width} we list values of $\sqrt{{\overline{s}}_{\Delta}}$ 
and the associated in-medium decay-pion momentum ${\overline{q}}_{\Delta\to N
\pi}$ for several representative values of the r.m.s. radius $R_{\Delta\Delta}
\equiv {\langle r_{\Delta\Delta}^2\rangle}^{1/2}$ of the bound $\Delta\Delta$ 
wavefunction, obtained from Eq.~(\ref{eq:av}) by using the equality sign in 
the uncertainty relationship $P_{\Delta\Delta}R_{\Delta\Delta}\geq 3/2$, in 
units of $\hbar=c=1$. Listed also are values of the in-medium single-$\Delta$ 
width ${\overline{\Gamma}}_{\Delta\to N\pi}$, obtained from the empirical 
$\Delta$-decay momentum dependence 
\begin{equation}
{\overline{\Gamma}}_{\Delta\to N\pi}({\overline{q}}_{\Delta\to N\pi}) = 
\gamma\,\frac{{\overline{q}}^3_{\Delta\to N\pi}}{q_0^2+{\overline{q}}^2_{
\Delta\to N\pi}}, 
\label{eq:gamma} 
\end{equation} 
with $\gamma=0.74$ and $q_0=159$~MeV~\cite{BCS17}. 
By relating ${\overline{q}}_{\Delta\to N\pi}$ in this expression to 
$\sqrt{{\overline{s}}_{\Delta}}$ of Eq.~(\ref{eq:s}) in the same way as 
in free space, it is implicitly assumed here that this empirical momentum 
dependence provides a good approximation also for off-shell $\Delta$s. 
Finally, The last column of the table lists values of ${\overline{\Gamma}}_{
\Delta\Delta\to NN\pi\pi}$ obtained by multiplying ${\overline{\Gamma}}_{
\Delta\to N\pi}$ by two, for the two $\Delta$s, while applying to one of them 
the isospin projection factor 2/3 introduced in the Gal-Garcilazo hadronic 
model~\cite{galgar13,galgar14} to satisfy the quantum statistics requirements 
in the leading final $NN\pi\pi$ decay channels. The large spread of 
${\overline{\Gamma}}_{\Delta\Delta\to NN\pi\pi}$ width values exhibited in the 
table, all of which are much smaller than the 162~MeV obtained by ignoring 
in Eq.~(\ref{eq:s}) the bound-state momentum distribution, demonstrates 
the importance of this momentum contribution. It is seen that a compact 
$d^{\ast}$(2380) with values of $R_{\Delta\Delta}$ between 0.6 to 0.8~fm is 
incompatible with the experimental value $\Gamma_{d^{\ast}}$(2380)=80$\pm
$10~MeV from WASA-at-COSY and SAID~\cite{wasa14} even upon adding a non-pionic 
partial width $\Gamma_{\Delta\Delta\to NN}\sim 10$~MeV~\cite{BCS17}. 
In particular, $R_{\Delta\Delta}$=0.76~fm from the quark-based model of 
Ref.~\cite{huang15}, as shown on the l.h.s. panel of Fig.~\ref{fig:wf}, 
leads to an unacceptably small value of about 47~MeV for the width.{
\footnote{This is an upper bound, given that the equality sign was used in 
the uncertainty relationship. Using an infinite square well of radius 1.43~fm 
that gives $R_{\Delta\Delta}$=0.76~fm in the g.s., one gets a value of 27~MeV 
instead of 47 MeV for ${\overline{\Gamma}}_{\Delta\Delta\to NN\pi\pi}$}.} 
This drastic effect of momentum dependence is missing in quark-based 
decay-width calculations of a single $\Delta\Delta$ configuration, e.g. 
Ref.~\cite{dong16}, which would underestimate considerably the 
$d^{\ast}$(2380) width once the momentum distribution of a tightly-bound 
and compact $\Delta\Delta$ is accounted for. 

The preceding discussion of the $d^{\ast}$(2380) width suggests that the 
quark-based model's finding of a tightly bound $\Delta\Delta$ $s$-wave 
configuration is in conflict with the observed width. Fortunately, the 
hadronic-basis calculations mentioned in the Introduction offer resolution 
of this insufficiency by adding to the tightly bound and sub-fm compact 
$\Delta\Delta$ component of the $d^{\ast}$(2380) dibaryon's wavefunction 
a $\pi N\Delta$ resonating component dominated asymptotically by a $p$-wave 
pion attached loosely to the near-threshold $N\Delta$ dibaryon ${\cal D}_{12}$ 
with size about 1.5--2~fm. Formally, one can recouple spins and 
isospins in this $\pi{\cal D}_{12}$ system, as demonstrated in the Appendix, 
so as to assume an extended $\Delta\Delta$-like object. This explains why 
the preceding discussion of $\Gamma_{d^{\ast}\to NN\pi\pi}$ in terms of 
a $\Delta\Delta$ constituent model required a size larger than provided by 
corresponding quark-based RGM calculations~\cite{dong16}. We recall that the 
$\pi N\Delta$ model~\cite{galgar13,galgar14} does reproduce the observed width 
of the $d^{\ast}$(2380) dibaryon resonance. 
The relevance of the ${\cal D}_{12}(2150)$ $N\Delta$ dibaryon to the physics 
of the $d^{\ast}$(2380) resonance is also demonstrated on the r.h.s. of 
Fig.~\ref{fig:wf} by showing a $d\pi$ invariant-mass distribution peaking 
near the $N\Delta$ threshold as deduced from the $pn\to d\pi^0\pi^0$ reaction 
by which the $d^{\ast}$(2380) was discovered~\cite{wasa11}. This peaking, 
essentially at the ${\cal D}_{12}(2150)$ mass value, suggests that the 
$\pi{\cal D}_{12}$ two-body channel plays an important role in the 
decay modes of the $d^{\ast}$(2380) dibaryon, as reflected in the calculation 
of Ref.~\cite{PK16} depicted in the figure. The width of this invariant-mass 
distribution, nevertheless, agrees roughly with $\Gamma_{d^{\ast}}$(2380)=80$
\pm$10~MeV irrespective of the underlying decay mechanism.  

To end this discussion of the two-channel structure of the $d^{\ast}$(2380) 
dibaryon resonance, we mention the ABC effect \cite{ABC60} which has been 
debated extensively in the context of the $d^{\ast}$(2380) dibaryon resonance 
\cite{clement17}. For a recent study see Ref.~\cite{BCS17}. Here, one observes 
a pronounced low-mass enhancement at $M_{\pi^0\pi^0}\sim 0.3$~GeV in the 
$\pi^0\pi^0$ invariant mass distribution of the $pn\to d\pi^0\pi^0$ fusion 
reaction at $\sqrt{s}=2.38$~GeV. Realizing that the decay pions from a $d^{
\ast}$(2380) compact $\Delta\Delta$ component have particularly low momenta, 
we compute $M_{\pi^0\pi^0}=314$~MeV by using the value ${\overline{q}}_{
\Delta\to N\pi}=113.6$~MeV/c, corresponding to $R_{\Delta\Delta}=0.76$~fm 
from the quark-based calculations of Ref.~\cite{huang15}. The ABC enhancement 
appears not to arise in the $pn\to pn\pi^0\pi^0$ non-fusion reaction, 
apparently because the outgoing quasi-free nucleons manage to affect the 
$\Delta\to N\pi$ decay spectra more readily than when bound in the deuteron. 
Furthermore, it was found in Ref.~\cite{BCS17} that to reproduce the shape 
of the $M_{\pi^0\pi^0}$ distribution relative to the ABC enhancement, 
a form factor of size approximately 2~fm is required. This would correspond 
in the present two-channel approach roughly to the size of the resonating 
$\pi{\cal D}_{12}$ component of the $d^{\ast}$(2380) dibaryon. 
More work is needed to substantiate these suggestions.

\section{$d^{\ast}$(2380) partial decay widths and branching ratios} 
\label{sec:BR} 

Here we evaluate the $d^{\ast}$(2380) partial decay widths and branching 
ratios (BR). Various pieces of experimental and theoretical input to the 
$d^{\ast}$(2380) production and decay data are incorporated in this 
evaluation as follows. 
\begin{enumerate} 
\item 
A value of $\Gamma_{\rm tot}^{d^{\ast}}=75$~MeV was adopted for the 
$d^{\ast}$(2380) total width to allow direct comparison with the analysis of 
Ref.~\cite{BCS15}. This value is close to $\Gamma_{\rm tot}^{d^{\ast}}=70$~MeV 
derived from the observed $pn\to d\pi^0\pi^0$ resonance shape \cite{wasa11}, 
and is within the range of values $\Gamma_{\rm tot}^{d^{\ast}}=80\pm 10$~MeV 
determined by the SAID analysis of the WASA-at-COSY recent measurements 
of polarized ${\vec n}p$ elastic scattering around the $d^{\ast}$(2380) 
resonance \cite{wasa14}. 
\item 
$NN$ partial decay widths between 9 to 11~MeV were used for $\Gamma_{NN}^{
d^{\ast}}$, corresponding to BR between 0.12 and 0.15, in agreement with 
$\Gamma_{NN}^{d^{\ast}}/\Gamma_{\rm tot}^{d^{\ast}}=0.12 \pm 0.03$ from the 
SAID determination \cite{wasa14} and with the value 0.15 extracted from the 
$pn$ $^3D_3$ Argand diagram shown in Fig.~\ref{fig:WASA}. The actual choice 
of $\Gamma_{NN}^{d^{\ast}}$ is described in item~5 below. 
\item 
A $d^{\ast}$(2380) resonance peak value of $\sigma(pn\to d^{\ast}\to d\pi^0
\pi^0)=240$~$\mu$b was assumed, following Ref.~\cite{BCS15}, to determine the 
product $\Gamma_{NN}^{d^{\ast}}\Gamma_{d\pi^0\pi^0}^{d^{\ast}}$, and hence 
the value of $\Gamma_{d\pi^0\pi^0}^{d^{\ast}}$. For $\Gamma_{d\pi^+\pi^-}^{d^{
\ast}}$ we multiplied $\Gamma_{d\pi^0\pi^0}^{d^{\ast}}$ by 1.83 \cite{dong16}, 
close to the pure isospin limit of 2, and followed the latter work also to 
obtain $\Gamma_{pn\pi^0\pi^0}^{d^{\ast}}$, $\Gamma_{pn\pi^0\pi^0}^{d^{
\ast}}=1.04\times\Gamma_{d\pi^0\pi^0}^{d^{\ast}}$ in rough agreement with 
Refs.~\cite{FW11,oset13}. The obtained value of $\Gamma_{pn\pi^0\pi^0}^{d^{
\ast}}$ was then multiplied by the same factor 1.83 as above to get the 
{\it isoscalar} part of $\Gamma_{pn\pi^+\pi^-}^{d^{\ast}}$. With these 
values, the summed isoscalar part of $\Gamma_{NN\pi\pi}^{d^{\ast}}$ amounts 
to $5.77\times\Gamma_{d\pi^0\pi^0}^{d^{\ast}}$. Note that nowhere in 
this derivation have we relied on the quark-based model work \cite{dong16} 
{\it total} decay width $\Gamma_{NN\pi\pi}^{d^{\ast}}$ with which, according 
to the discussion in Sect.~\ref{sec:compact}, we disagree. 
\item 
To get the {\it isovector} part of $\Gamma_{NN\pi\pi}^{d^{\ast}}$, which is 
not related directly by isospin to the isoscalar part, the summed isoscalar 
part of $\Gamma_{NN\pi\pi}^{d^{\ast}}$, plus $\Gamma_{NN\pi}^{d^{\ast}}$, plus 
$\Gamma_{NN}^{d^{\ast}}$, were subtracted from $\Gamma_{\rm tot}^{d^{\ast}}$. 
As for $\Gamma_{NN\pi}^{d^{\ast}}$, it was extracted in a model-dependent 
way discussed below from the $\pi{\cal D}_{12}$ component of $d^{\ast}$(2380) 
by using a BR $\Gamma_{NN}^{{\cal D}_{12}}/\Gamma_{\rm tot}^{{\cal D}_{12}}
\approx 0.18$, taken from the Argand diagram of the $NN$ $^1D_2$ partial 
wave in the SAID SP07 fit \cite{SP07}. 
\item 
The dependence of the isovector part of $\Gamma_{NN\pi\pi}^{d^{\ast}}$ on 
$\Gamma_{NN}^{d^{\ast}}$ was used to choose a value for $\Gamma_{NN}^{d^{
\ast}}$ (see item~2 above) so as to reproduce the $d^{\ast}$(2380) resonance 
peak value of $\sigma(pn\to d^{\ast}\to pp\pi^-\pi^0)\approx 100\pm 
10$~$\mu$b~\cite{wasa13c,BCS15}. 
\end{enumerate} 

The $d^{\ast}$(2380) partial decay widths (in MeV) and the corresponding 
BR (in percents) derived using these specifications are listed 
in Table~\ref{tab:BR} within (i) a pure $\Delta\Delta$ model ($\alpha=1$), 
(ii) a pure $\pi{\cal D}_{12}$ model ($\alpha=0$), and (iii) within a 
$d^{\ast}$(2380) $\Delta\Delta$--$\pi{\cal D}_{12}$ mixing model ($\alpha=
\frac{5}{7}$). The $\Delta\Delta$ decay fraction $\alpha$ is defined by 
Eq.~(\ref{eq:alpha}) below. In this mixing model, the $d^{\ast}$(2380) 
resonance consists of a superposition of inner $\Delta\Delta$ and outer 
$\pi{\cal D}_{12}(2150)$ components. The $NN\pi\pi$ decays from these two 
components involve quite different portions of phase space, and their 
associated widths add up incoherently. For the $NN\pi\pi$ decay width of 
the compact $\Delta\Delta$ component we chose a value of $\Gamma_{<}=44$~MeV, 
in between the values listed in Table~\ref{tab:width} for $R_{\Delta\Delta}
=0.7$ and 0.8~fm. A corresponding value of $\Gamma_{>}=100$~MeV, inspired by 
the ${\cal D}_{12}(2150)$ total width of 120~MeV derived by solving the 
appropriate $\pi NN$ Faddeev equations~\cite{galgar13,galgar14}, from which 
we subtracted $\approx$20~MeV for the $NN$ decay mode, was chosen for the 
$\pi{\cal D}_{12}(2150)$ asymptotic component. Assigning $NN\pi\pi$ decay 
fractions $\alpha$ and $1-\alpha$, respectively, we solved the equation 
\begin{equation} 
\alpha \Gamma_{<} + (1-\alpha) \Gamma_{>} = \Gamma_{NN\pi\pi}^{d^{\ast}}. 
\label{eq:alpha} 
\end{equation} 
With $\Gamma_{NN}^{d^{\ast}}\approx 10$~MeV, and expecting $\Gamma_{NN\pi}^{
d^{\ast}}\approx 5$~MeV, we estimate $\Gamma_{NN\pi\pi}^{d^{\ast}}=60$~MeV. 
The value of $\alpha$ that solves this equation is $\alpha=\frac{5}{7}$. 
Future hadronic calculations should tell how good this representative value of 
$\alpha$ is. The partial decay widths and BR resulting in this mixing version 
are listed in Table~\ref{tab:BR} under the heading $\alpha=\frac{5}{7}$. 

\begin{table}[hbt] 
\begin{center} 
\caption{Partial widths ($\Gamma_f^{d^{\ast}}$ in MeV) and branching 
ratios (BR in percents) for $d^{\ast}(2380)$ decays, calculated in 
a $\Delta\Delta$--$\pi{\cal D}_{12}$ coupled channels scheme specified 
by values for the mixing parameter $\alpha$ (see text) and for 
$\Gamma_{NN}^{d^{\ast}}$ (row before last). The total width 
$\Gamma_{\rm tot}^{d^{\ast}}=75$~MeV, a peak cross section value 
$\sigma(pn\to d^{\ast}\to d\pi^0\pi^0)=240$~$\mu$b \cite{BCS15} and 
a ratio $\Gamma_{NN}^{{\cal D}_{12}}/\Gamma_{\rm tot}^{{\cal D}_{12}}=0.18$ 
\cite{SP07} are held fixed.} 
\begin{tabular}{cccccccc} 
\hline 
final & \multicolumn{2}{c}{$\Delta\Delta$ ($\alpha=1$)} & \multicolumn{2}{c}
{$\pi{\cal D}_{12}$ ($\alpha=0$)} & \multicolumn{2}{c}{mixed ($\alpha=
\frac{5}{7}$)} & exp.\cite{BCS15}  \\
state & $\Gamma_f^{d^{\ast}}$ & BR & $\Gamma_f^{d^{\ast}}$ & BR & 
$\Gamma_f^{d^{\ast}}$ & BR & BR  \\ 
\hline 
$d\pi^0\pi^0$ & 9.3 & 12.4 & 7.6 & 10.1 & 8.4 & 11.2 & 14(1) \\ 
$d\pi^+\pi^-$ & 17.0 & 22.7 & 14.0 & 18.6 & 15.3 & 20.4 & 23(2) \\ 
$pn\pi^0\pi^0$ & 9.7 & 12.9 & 7.9 & 10.5 & 8.7 & 11.6 & 12(2) \\ 
$pn\pi^+\pi^-$ & 21.7 & 28.9 & 17.2 & 22.9 & 19.3 & 25.8 & 30(5) \\ 
$pp\pi^-\pi^0$ & 4.15 & 5.55 & 2.9 & 3.9 & 3.55 & 4.7 & 6(1) \\ 
$nn\pi^+\pi^0$ & 4.15 & 5.55 & 2.9 & 3.9 & 3.55 & 4.7 & 6(1) \\ 
$NN\pi$ & -- & -- & 11.5 & 15.4 & 6.2 & 8.3 & -- \\ 
$NN$ & 9 & 12 & 11 & 14.7 & 10 & 13.3 & 12(3) \\ 
total & 75 & 100 & 75 & 100 & 75 & 100 & 103(15) \\ 
\hline 
\end{tabular}  
\label{tab:BR} 
\end{center} 
\end{table} 

Comparing the BR obtained in the three model versions specified by their value 
of the $\Delta\Delta$ fraction $\alpha$ with those derived from experiment 
in Ref.~\cite{BCS15} and listed in the last column of Table~\ref{tab:BR}, 
one notes the similarity between the BR obtained in a purely $\Delta\Delta$ 
model ($\alpha=1$) and those derived from experiment. In fact, this similarity 
is somewhat fortuitous because all three model versions were designed to 
reproduce input values of the $d^{\ast}$ peak cross sections: $\sigma(pn\to 
d^{\ast}\to d\pi^0\pi^0)=240$~$\mu$b and $\sigma(pn\to d^{\ast}\to pp\pi^-
\pi^0)\approx 100\pm 10$~$\mu$b \cite{BCS15}, thereby agreeing also for the 
rest of the $pn\to d^{\ast}\to NN\pi\pi$ cross sections. The three model 
versions figuring in Table~\ref{tab:BR} differ essentially {\it only} in 
their $NN\pi$ BR which in the purely $\Delta\Delta$ model ($\alpha=1$) is 
close to zero~\cite{dong17}. The $NN\pi$ partial decay width and 
BR listed for the purely $\pi{\cal D}_{12}$ model ($\alpha=0$) were normalized 
to a total $d^{\ast}$ pionic width of 75$-$11=64~MeV. The relatively 
high value of $\approx$15\% for the obtained BR is excluded by a recent 
determination of a $\lesssim 9\%$ upper limit \cite{wasa17}. In contrast, 
a value of the $NN\pi$ BR smaller by almost a factor of two was obtained, 
by applying the $\pi{\cal D}_{12}$ decay fraction $(1-\alpha)$ to the $NN$ 
partial decay width $\Gamma_{NN}^{{\cal D}_{12}}=0.18\times 120$~MeV, 
in the specific mixing model version listing in Table~\ref{tab:BR}. 

How robust are the BR results shown for the $\Delta\Delta$--$\pi{\cal D}_{12}$ 
coupled channels scheme in Table~\ref{tab:BR}? The listed BR are based on 
assuming a value $\Gamma_{NN}^{d^{\ast}}=10$~MeV. A $\pm$10\% variation 
of this value results in $\approx\pm$50\% variation in the cross section 
$\sigma(pn\to d^{\ast}\to pp\pi^-\pi^0)$ away from its initially assumed value 
which can be restored by a $\pm$10\% variation in $\sigma(pn\to d^{\ast}\to 
d\pi^0\pi^0)$ away from its initially assumed value. We conclude that the 
partial decay widths and BR listed in Table~\ref{tab:BR} for a mixing 
parameter $\alpha=\frac{5}{7}$ have uncertainties of up to 10\%, except for 
those for the $NN\pi$ decay mode which depend only on the assumed value of 
$\alpha$. 

Next we allow $\alpha$ to vary by replacing the value of $\Gamma_{<}=44$~MeV 
that served as input through Eq.~(\ref{eq:alpha}) to derive the value 
of $\alpha=\frac{5}{7}$ in use in Table~\ref{tab:BR} by representative 
neighboring values 40 and 50~MeV. The resulting values of $\alpha$ are 
$\alpha=\frac{2}{3}$ and $\frac{4}{5}$, respectively, leading to the following 
uncertainty estimate for the $NN\pi$ decay mode:      
\begin{equation} 
\sigma(pn\to d^{\ast}\to NN\pi)=178^{+29}_{-55}~\mu{\rm b}, \,\,\,\,\, 
\Gamma_{NN\pi}^{d^{\ast}}\approx 6.2^{+1.0}_{-1.9}~{\rm MeV}, \,\,\,\,\, 
\frac{\Gamma_{NN\pi}^{d^{\ast}}}{\Gamma_{\rm tot}^{d^{\ast}}}\approx 
8.3^{+1.3}_{-2.5}~\%. 
\label{eq:BR} 
\end{equation} 
Note that in addition to the $NN\pi$ partial decay width 
$\Gamma_{NN\pi}^{d^{\ast}}$ and branching ratio $\Gamma_{NN\pi}^{d^{\ast}}/
\Gamma_{\rm tot}^{d^{\ast}}$, with central values as given already in 
Table~\ref{tab:BR}, we have also provided here a cross-section estimate for 
$\sigma(pn\to d^{\ast}\to NN\pi)$, with estimated uncertainties, to compare 
directly with the experimental upper limit of 180~$\mu$b~\cite{wasa17}. 
Given these uncertainties, the $NN\pi$ production cross section could be 
as low as $\sim 120~\mu$b, comfortably below the reported upper limit.

\section{Conclusion} 
\label{sec:sum} 

The $d^{\ast}$(2380) is the most promising dibaryon candidate at present, 
supported by systematic studies of its production and decay in recent 
WASA-at-COSY experiments \cite{clement17}. In most theoretical works, 
beginning with the 1964 Dyson-Xuong prediction \cite{dyson64}, 
it is assigned as a $\Delta\Delta$ quasibound state. Given the small width 
$\Gamma_{d^{\ast}(2380)}=80\pm 10$~MeV with respect to twice the width of a 
free-space $\Delta$, $\Gamma_{\Delta}\approx 115$~MeV, its location far from 
thresholds makes it easier to discard a possible underlying threshold effect. 
However, as argued in this work, the observed small width is much larger than 
what two {\it deeply bound} $\Delta$ baryons can yield upon decay. The $d^{
\ast}$(2380) therefore cannot be described exclusively by a $\Delta\Delta$ 
component. A complementary quasi two-body component is offered in the $\pi 
N\Delta$ three-body hadronic model of Refs.~\cite{galgar13,galgar14} by 
a $\pi{\cal D}_{12}$ channel, in which the $d^{\ast}$(2380) resonates. 
The ${\cal D}_{12}$ dibaryon stands here for the $I(J^P)=1(2^+)$ $N\Delta$ 
near-threshold system that might or might not possess a quasibound state 
$S$-matrix pole. It is a loose system of size typically 1.5--2~fm, as opposed 
to the compact $\Delta\Delta$ component of size 0.5--1~fm. It was pointed 
out how the ABC low-mass enhancement in the $\pi^0\pi^0$ invariant mass 
distribution of the $pn\to d\pi^0\pi^0$ fusion reaction at $\sqrt{s}=2.38$~GeV 
might be associated with the small size of the $\Delta\Delta$ component. 
Furthermore, we have shown how to determine the relative weight of these 
two components by fitting to the total $d^{\ast}$(2380) width, thereby 
deriving $d^{\ast}$ partial decay widths and branching ratios that agree 
with experiment \cite{BCS15}. A new element in the present derivation 
is the ability, through the $\pi{\cal D}_{12}$ channel, to evaluate the 
$d^{\ast}\to NN\pi$ decay width and BR. Our prediction is for a BR of 
order 8\%, considerably higher than that obtained for a quark-based 
purely $\Delta\Delta$ configuration~\cite{dong17}, but consistently with 
an upper limit of $\lesssim 9\%$ determined recently by the WASA-at-COSY 
collaboration~\cite{wasa17}. A precise measurement of this decay 
width and BR will provide a valuable constraint on the 
$\pi{\cal D}_{12}$--$\Delta\Delta$ mixing parameter.

\section*{Acknowledgments} 
Special thanks are due to Heinz Clement for stimulating exchanges on 
the physics of dibaryons \cite{clement17}, and to Maria Platonova on 
the $\pi{\cal D}_{12}$ interpretation of the $d^{\ast}$(2380) \cite{PK16}.

\section*{Appendix: Equivalence of isospin bases} 

Starting with a ${\cal D}_{12}\otimes\pi$ structure of $d^{\ast}$(2380), we 
recall that ${\cal D}_{12}$ stands for a near-threshold $\Delta N$ dibaryon, 
where $\Delta$ is a $P_{33}$ $N\pi$ resonance, and recouple in isospace: 
\begin{equation} 
\left\{[(N_1\otimes\pi_1)_{\frac{3}{2}}\otimes N_2]_{I_{{\cal D}_{12}}}  
\otimes\pi_2\right\}_{I=0}\to \left\{[(N_1\otimes N_2)_{I_{NN}}\otimes 
\pi_1 ]_{I_{{\cal D}_{12}}} \otimes \pi_2\right\}_{I=0} , 
\label{eq:iso1} 
\end{equation} 
using the $6j$ orthogonal transformation with elements given by 
\begin{equation} 
(-1)^{I_{NN}+1}\sqrt{4(2I_{NN}+1)}\left\{\begin{array}{ccc}N_1&N_2&I_{NN} \\ 
I_{{\cal D}_{12}}&\pi_1&\frac{3}{2} \\ 
\end{array}\right\},  
\label{eq:iso2} 
\end{equation} 
where $N_1=N_2=\frac{1}{2}$ and $\pi_1=\pi_2=1$ denote the nucleons 
and pions isospins, respectively, and $I_{{\cal D}_{12}}=1$. 
The square of the $I_{NN}=0$ element is 2/3, and that of the $I_{NN}=1$ 
element is 1/3. The $I_{NN}=0$ projection factor 2/3 is the same as that 
considered in the Gal-Garcilazo hadronic model for the pionic decay modes 
of the $\Delta\Delta$ component. 
Note that the $NN\pi$ Faddeev calculation of ${\cal D}_{12}$ in 
Refs.~\cite{galgar13,galgar14} was based on a single $I_{NN}=0$, $S_{NN}=1$ 
$s$-wave configuration, thereby justifying the $I_{NN}=0$ projection applied 
here. Note also that with $S_{NN}=1$ and two $p$-wave pions, the angular 
momentum coupling needed for the $3^+$ $d^{\ast}$(2380) is unique, with 
all individual components parallel to each other. Therefore we need to 
focus just on the isospin recoupling. Proceeding to recouple the state on 
the r.h.s. of Eq.~(\ref{eq:iso1}),  
\begin{equation} 
\left\{[(N_1\otimes N_2)_{I_{NN}}\otimes \pi_1 ]_{I_{{\cal D}_{12}}}\otimes 
\pi_2\right\}_{I=0}\to[(N_1\otimes N_2)_{I_{NN}}\otimes (\pi_1 \otimes 
\pi_2)_{I_{\pi\pi}}]_{I=0}, 
\label{eq:iso3} 
\end{equation} 
where $I_{\pi\pi}=I_{NN}$, we get $6j$ transformation elements identically 1,  
\begin{equation} 
(-1)^{I_{NN}}\sqrt{3(2I_{NN}+1)}\left\{\begin{array}{ccc}I_{NN}&\pi_1&
I_{{\cal D}_{12}} \\ \pi_2&0&I_{\pi\pi} \\ \end{array}\right\} = 1 , 
\label{eq:iso4} 
\end{equation} 
irrespective of the value of $I_{NN}=I_{\pi\pi}$. 
In Refs.~\cite{galgar13,galgar14} we got the state on the r.h.s. 
of Eq.~(\ref{eq:iso3}) by recoupling directly from a $\Delta\Delta$ 
configuration, 
\begin{equation} 
[(N_1\otimes \pi_1)_{\frac{3}{2}}\otimes 
(N_2\otimes \pi_2)_{\frac{3}{2}}]_{I=0} \to 
[(N_1\otimes N_2)_{I_{NN}}\otimes 
(\pi_1 \otimes \pi_2)_{I_{\pi\pi}}]_{I=0},   
\label{eq:iso5} 
\end{equation} 
using the $9j$ transformation 
\begin{equation} 
4\sqrt{(2I_{NN}+1)(2I_{\pi\pi}+1)}\left\{\begin{array}{ccc}N_1&\pi_1&
\frac{3}{2} \\ N_2&\pi_2&\frac{3}{2} \\ I_{NN}&I_{\pi\pi}&0 \\ 
\end{array}\right\} 
\label{eq:iso6} 
\end{equation} 
which yields precisely the same $6j$ transformation elements as 
in Eq.~(\ref{eq:iso2}). This establishes the equivalence of the 
${\cal D}_{12}\pi$ and $\Delta\Delta$ bases as far as the calculation of 
$\Gamma_{\Delta\Delta\to NN\pi\pi}$ branching ratios in section~\ref{sec:BR} 
is concerned.

\end{document}